# Novel self-locked architecture for ultra-stable micro-ring resonator based optical frequency combs

Alessia Pasquazi,[1] Lucia Caspani,[1] Marco Peccianti,[2,*] Matteo Clerici,[1,3] Marcello Ferrera,[1] Luca Razzari,[1] David Duchesne,[4] Brent E. Little,[5] Sai T. Chu,[6] David J. Moss,[7,8] and Roberto Morandotti[1]

[1]*INRS - Énergie, Matériaux et Télécommunications, 1650 Blvd Lionel Boulet, Varennes (Québec), J3X1S2, Canada.*
[2]*Institute for Complex Systems, CNR, Via dei Taurini 19, 00185 Roma, Italy.*
[3]*School of Engineering and Physical Sciences, Heriot-Watt University, SUPA, Edinburgh EH14 4AS, UK.*
[4]*Massachusetts Institute of Technology, Cambridge, Massachusetts 02139, USA*
[5]*Infinera Ltd, 169 Java Drive, Sunnyvale, California 94089, USA.*
[6]*City University of Hong Kong, Department of Physics and Material Science, Tat Chee Avenue, Hong Kong, China.*
[7]*CUDOS, the Institute of Photonics and Optical Science (IPOS), School of Physics, University of Sydney, Sydney, NSW 2006, Australia.*
[8]*Present Address: School of Electrical and Computer Engineering, RMIT University, Melbourne, Vic. Australia 3001*
*m.peccianti@gmail.com

**Abstract:** We report a novel geometry for OPOs based on nonlinear microcavity resonators. This approach relies on a self-locked scheme that enables OPO emission without the need for thermal locking of the pump laser to the microcavity resonance. By exploiting a CMOS-compatible microring resonator, we achieve oscillation with a complete absence of shutting down, or self-terminating behavior, a very common occurrence in externally pumped OPOs. Further, this scheme consistently produces very wide bandwidth (>300nm, limited by our experimental set-up) combs that oscillate at a spacing of the FSR of the micro cavity resonance.

## 1. Introduction

Optical frequency combs (OFCs) are optical radiation patterns characterized by a spectrum of discrete equally spaced frequencies or *modes*. Amongst many other things, they are potentially highly accurate optical clocks, and highly stable, broadband OFCs are having an enormous impact on metrology and spectroscopy. Indeed, they have recently enabled the measurement of physical constants with an unprecedented degree of accuracy and have thus facilitated new observations in a wide range of disciplines from astronomy to geology and biology [1-5]. Moreover, there is a common belief that OFCs will play a key role in optical signal processing - the full control of the phase of an OFC will enable the synthesis and measurement of arbitrary optical signals [6-7], directly impacting diverse fields ranging from telecommunications, to the emerging ultrafast micro-chip computing. The recent realization of multiple wavelength sources in monolithic form [9-12] represents a fundamental advance in this direction. These sources are typically achieved by resonantly coupling an *external* laser into a nonlinear micro-cavity. A nonlinear field-matter interaction, namely four-wave mixing (FWM), induces optical parametric oscillation (OPO), which subsequently generates a comb of equally spaced lines. Since the first breakthroughs based on micro-spheres and micro-toroids [7-12], these sources have been demonstrated in planar glass technologies compatible with integrated CMOS (complementary metal-oxide semiconductor) platforms [13-15], opening up the unique possibility of using these devices as light sources in microchips. These sources are in general characterized by a comb teeth spacing ranging from 80GHz to more than a THz, making them particularly appealing for telecom and optical clocks applications [10]. Several routes to comb formation have been observed [15-18], leading to either chaotic or coherent regimes. The optical pulses generated in the coherent regime have sometimes been classified as optical cavity solitons since the microcavity OPO can be effectively described by the standard nonlinear Schrödinger equation (NLSE) [19-22]. A critical aspect regarding this point is that the generation of optical solitons in a dissipation-free system such as those described by the NLSE is strongly dependent on the initial state of the system and the launch conditions of the pulse [22]. Hence, controlling the coupling of the external pump laser to the microcavity is essential for achieving and maintaining the coherence of the emitted signal.

A common issue faced by *all* microresonator based OPOs is the instability of the locking between the external pump laser and the resonances of the microcavity. Due to the small mode volume of the microcavities, the absorbed pump power heats the cavity and induces a thermal drift of the optical resonances [23]. *Soft-thermal locking* can be exploited to solve this issue by slowly tuning the pump wavelength, following the thermal drift of the cavity resonances, until locking is finally achieved. However, this method is ineffective against external slow temperature variations and pump power instabilities, resulting in an unstable regime where the OPO often fails to maintain oscillation. Most importantly, it fixes the coupled power to the excitation frequency [23], effectively reducing the degrees of freedom for realizing the control necessary to manage the coherence of the oscillation.

We recently introduced a novel, passive mode-locking scheme that addresses some of these issues [24]. In such a scheme, termed Filter Driven Four Wave Mixing (FD-FWM), a nonlinear high Q factor microresonator is nested in a fiber ring cavity. The external main fiber ring cavity, with a much more finely spaced FSR, features an active element that provides gain, while the micro-resonator is responsible for mode-locking via FWM. This system is dissipative, and hence is qualitatively very different to those described by the NLSE. In dissipative systems passive mode-locking arises from the interplay between dispersion, nonlinearity and gain / loss, and so the system naturally relaxes towards a stable, robust, dissipative soliton attractor [25]. With this scheme we demonstrated that a stable, fully mode-locked pulse train can be generated at ultra-high repetition rates (>200GHz) directly on-chip, sustained by FWM in a microresonator.

In addition to providing the optical gain, the active cavity also determines the modes of oscillation - the microresonator selects and mixes only the modes falling inside its own resonances. The stable regime occurs when a single mode of the main fiber cavity oscillates within each micro-cavity resonance. Further, more complex stable states of oscillation can be achieved, for example when *two* fiber loop modes oscillate simultaneously within each microcavity mode. In this case the system can deliver a stable radio frequency beat tone superimposed on the high repetition rate optical pulse train [26]. For higher numbers of oscillating modes within each microcavity resonance, operation typically becomes chaotic [24, 26].

In this paper, we present a new approach to generating OFCs that intrinsically achieves operation that is free from thermally induced self-termination, a condition that conventional externally pumped schemes are notorious for exhibiting and which often can only be addressed through the use of elaborate external feedback or locking mechanisms. Following our preliminary results [27], the approach we introduce here uses a self-locked scheme to achieve operation with a complete absence of shut down (self-termination). Further, this scheme also consistently produces very wide bandwidth (>300nm full observable bandwidth) combs that oscillate at the FSR spacing of the microcavity. We achieve this by adapting the "nested cavity" approach, mentioned above, to realize stable mode locking [24] by *tailoring* the laser gain profile that each mode experiences through the use of an in-loop fiber bandpass filter. Hence, only the main cavity modes lying within *one* particular resonance of the micro-cavity are amplified above threshold. These modes subsequently act as a pump that induces optical parametric oscillation in the same manner as standard microresonator OFCs. Here, however, the pump is not externally applied but self-oscillates within a resonance of the micro-cavity, and so the system is stabilized via an intrinsic, i.e. not extrinsic, feedback mechanism against thermal or mechanical fluctuations in the ring parameters. In our scheme the power is effectively a free parameter, provided that an appropriate level of gain is present in the fiber cavity, since the laser oscillation is restricted by the filter to lie within the same ring resonance. Such a feature is independent of the circulating optical power [23]. Indeed, we were able to consistently generate a 300nm wide bandwidth OPO, independently of the precise central pump wavelength within the gain bandwidth of the amplifier, i.e. a variation of almost 30nm (subject to meeting the resonant conditions for the ring). Further, we were also able to achieve the additional feature of producing a narrowly spaced OFC, at the FSR of the microring, rather than the much wider spacings often observed [14] at the peak MI gain wavelength. Thus, for all of these reasons we believe that this scheme will potentially play an important role in the quest to achieve ultimate control of the stability and coherence of microcavity based OFCs, rendering them much more robust and reliable for practical applications.

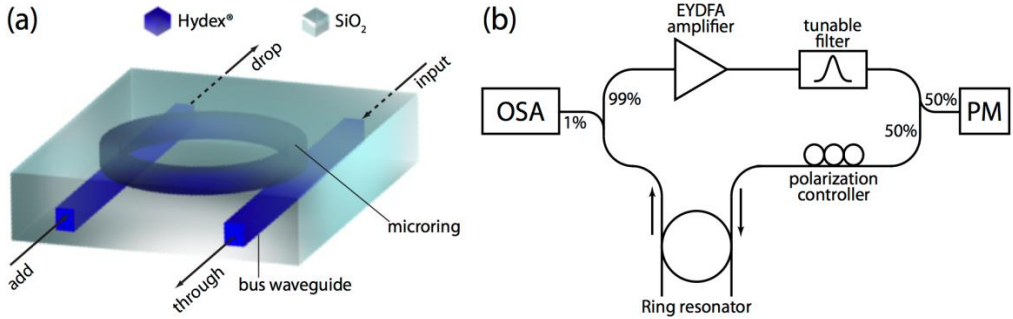

Fig 1. (a) Schematic of the 4-port vertically coupled microring resonator. The ring has a 135μm radius, corresponding to a FSR of 200.7GHz (≃1.6nm), and the waveguide section is 1.45×1.50 μm$^2$. The waveguide core is low-loss, high-index (n=1.7) doped silica glass, buried within a SiO$_2$ cladding. The advantages of this platform reside in its negligible linear and nonlinear losses [28-29] as well as in a nonlinear parameter as high as $\gamma \sim 0.22$W$^{-1}$m$^{-1}$ [28]. (b) Layout of the experimental setup: the output of the microring is almost all re-injected in the input port after proper amplification and filtering. OSA: optical spectrum analyzer; PM: power meter. The pump central wavelength is selected by means of the tunable filter (0.6nm FWHM).

## 2. Experiment

Figure 1 shows the laser configuration along with a schematic of the microring resonator - a high Q microring with a 160MHz linewidth and a FSR=200.7 GHz, (Q=$1.2\times10^6$) [28-30]. The waveguide core is low-loss, high-index (n=1.7) doped silica glass, buried within a $SiO_2$ cladding [28]. The device was fiber pigtailed to a standard SMF fiber with a typical coupling loss of 1.5dB/facet. The waveguide cross section is 1.45 µm x 1.50 µm while the ring diameter is 270µm. This platform shows negligible linear (< 6dB/meter) and nonlinear losses as well as a nonlinear parameter of $\gamma \sim 220 W^{-1}km^{-1}$ [28-30]. The microring resonator was embedded in a fiber loop cavity containing an erbium-ytterbium doped fiber amplifier (EYDFA), acting as the gain medium. The loop also contained a tunable bandpass filter (0.6 nm bandwidth), which allowed both the selection and tuning of the oscillating wavelength $\lambda_0$ within the bandwidth of a single microring resonance, in contrast to the FD-FWM mode locked laser described above. The cavity also contained an optical isolator and a polarization controller. The total main cavity length yielded an FSR of about 4.8MHz. A configuration with a shorter erbium doped amplifier (EDFA), employing an output beam sampler embedded in a free space delay line, was also implemented with an FSR of 64MHz. This system, however, had lower available optical gain than the longer laser. The polarization controller allowed us to properly adjust the polarization to either TE or TM. The output was then monitored by an optical spectrum analyzer via the optical output coupler which was a 1% - 99% tap filter.

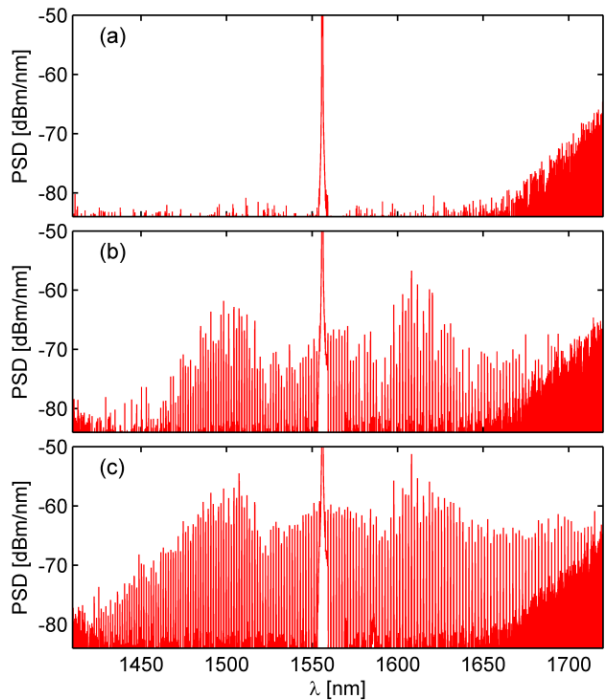

Fig. 2. Microring output spectrum (power spectral density, PSD) recorded with the OSA. These results are obtained with a pump wavelength $\lambda_p$=1556nm (corresponding to a TE mode) (a) Lasing of the pump at 1mW (power in the microring). No OPO is measured. Note that our OSA has an increasing noise at wavelength longer than 1650nm. (b) Lasing at 7mW: the OPO is detected (c) Lasing at 10mW. A >300nm bandwidth OPO is obtained. (Note that the pump peak is off-scale).Please note that the optical power measured by the OSA is the 1% of the optical power generated by the ring resonator.

## 3. Large bandwidth OPO generation

Figures 2(a)-2(c) summarize the results of the OFC. The oscillation of the pump pulse was achieved via an EYDFA in order to obtain 1mW to 10mW of average optical power at the input of the microring with the pump centered at 1556nm. The optical spectrum (recorded with an OSA), shown in Fig. 2(c) displays a comb-like structure spaced by the FSR of the microring (200.7 GHz) extending over a total bandwidth exceeding 300nm. Further, we always observed an OFC output operating on all the lines of the microresonator, an important characteristic highly useful for many applications such as spectroscopy. This is in stark contrast with the conventional approach of resonantly coupling an external laser, where achieving an OFC spaced as finely as the FSR is a significant challenge [31]. To illustrate this point, we show an OFC generated by coupling an external CW pump laser (linewidth < 300kHz) into the micro-resonator (Fig. 3(a) and 3(b)). As previously reported [14], hyper-parametric oscillation can be observed, with spectral features compatible to those of "Type I" coherent OPOs [Fig. 3(a)] (see [15] for details), showing the characteristic pyramid spectral shape generated by the pump. With the same pumping scheme, a set of pyramid shaped mini-combs, each generated by one line of a hyper-parametric, pyramid-shaped spectrum could be obtained [see Fig. 3(b)]. This regime has been referred to as a "Type II" comb [15].

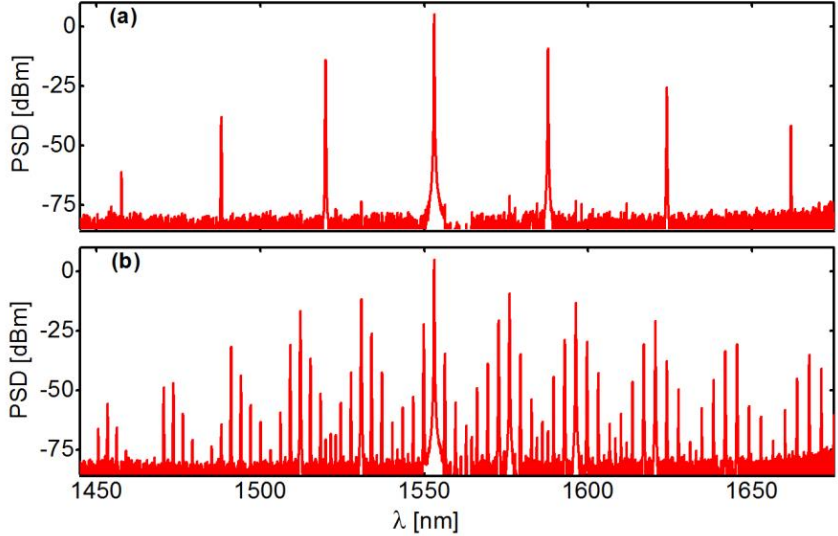

Fig. 3. Hyperparametric oscillation obtained by coupling an external laser with an input pump power (inside the bus waveguide) of 100mW and a linewidth of 150kHz centered at 1553.145 nm (a) The classical shape of coherent type I generation as defined in Ref [15] (b) Hyperparametric oscillation seeding secondary combs, characteristic of incoherent type II generation.

One of the most relevant features of our approach is the ability to sustain continuous operation virtually indefinitely. We observed no self-termination over the entire course of our experiments (several consecutive days) without the use of thermal or mechanical stabilization methods. Furthermore, the generated OFC is almost independent of the central pump wavelength, as is clearly seen in Fig. 4(a)-4(c) which shows the different spectra obtained by tuning the transmitted $\lambda_0$ wavelength of the intra-cavity filter, i.e. tuning the OPO pump wavelength. An OFC bandwidth substantially exceeding 300 nm was universally observed while tuning the pump wavelength over the entire 30 nm gain bandwidth of the EYDFA, independent of the coupled power. This highlights the robustness and versatility of our approach, since thermal locking intrinsically links the coupled power with the excitation wavelength. It is interesting to note that the spectral features of the comb are not particularly affected by the spectral position of the pump. However, we did observe that the OFC spectrum generated around the pump is flatter when the pump is coupled close to the zero dispersion point (e.g. Fig. 4 (c) where the pump oscillates at 1569nm).

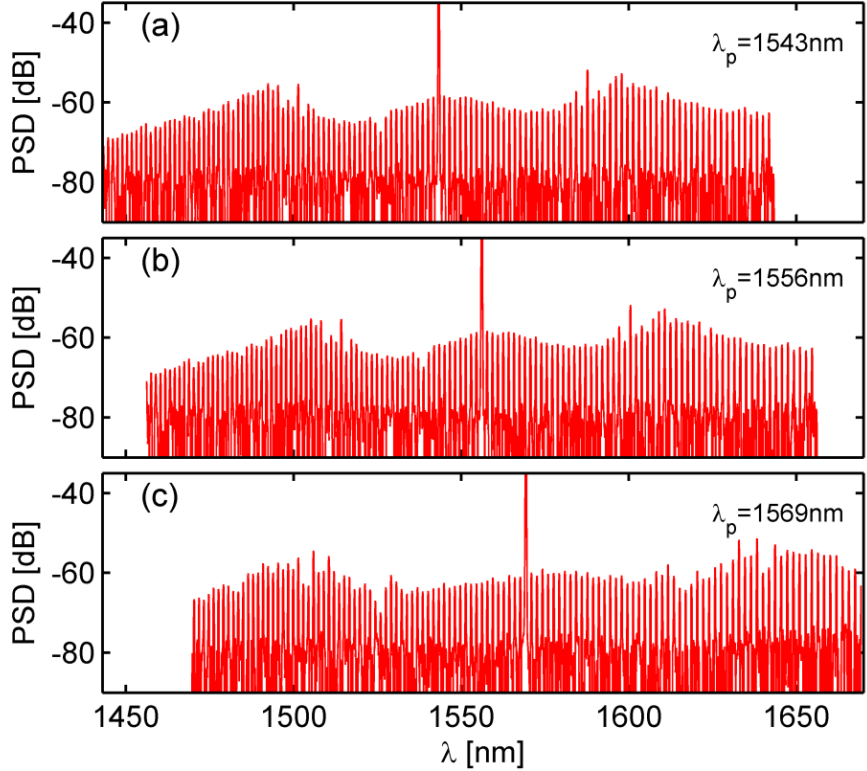

Fig. 4. Microring output spectrum recorded with the OSA for a pump power of ~30mW. (a) pump oscillating at 1543nm (b) pump oscillating at 1550n m (c) pump oscillating at 1569nm,

## 4. Discussion

This work clearly demonstrates that an efficient OPO, operating over a bandwidth exceeding 300nm, can be achieved using a "self-locked" scheme within a nested cavity configuration. We observed stable behavior, with resonances at different central frequencies and for different pumping regimes with a variation on the order of 0.67pm/mW power in cavity. However, in stable operating conditions we could not detect a shift of the central frequency with our OSA system with a resolution of 10pm.

While this new self-locked approach exhibits significant differences to the FD-FWM mode-locked laser [24] presented previously, it is nonetheless similar in spirit and hence displays some similar characteristics. In particular, the FD-FWM laser achieved extremely high optical stability in the RF regime by employing a very short cavity length laser in order to spread the main cavity modes out so that only a single longitudinal mode oscillated within each microcavity resonance, thereby eliminating the well-known problem of "supermode instability". Similarly, for the self-locked approach, although we restricted the pump wavelength to lie within a single microcavity, supermode instability was still present since more than a single main cavity mode could lase within the selected single ring resonance. Thus, while the overall operation of the laser never exhibited self-shutdown, it was still susceptible to RF oscillations, similar to [26]. Consequently, a low-frequency amplitude oscillation was observed in the long cavity (FSR=4.8MHz) output. Figures 5 (a)-5(f) show the RF spectrum of the pump at different input powers, for the cases shown in Fig. 2. The modes of the main cavity, oscillating at a frequency spacing of 4.8MHz are clearly visible.

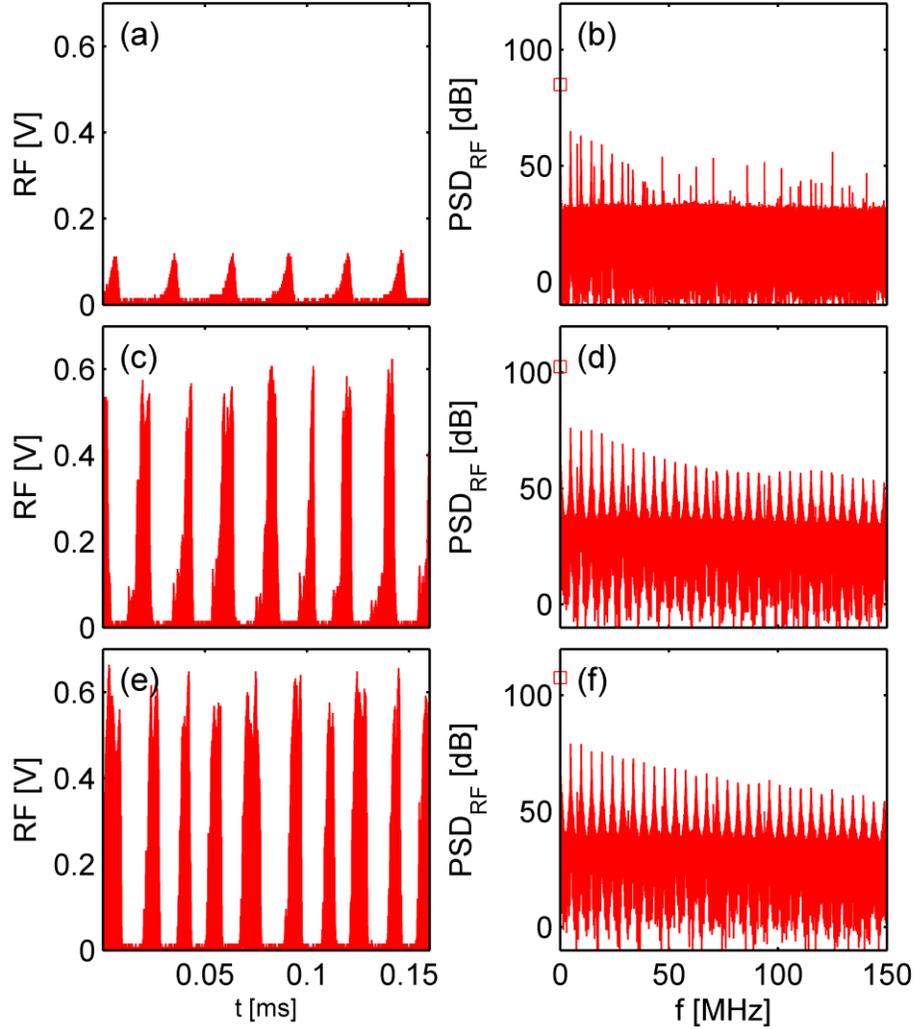

Fig.5. Microring output intensity measured with a fast photodetector. These results are obtained with a pump wavelength $\lambda_p$=1556nm. (corresponding to a TE mode) (a,b) Lasing of the pump at 1mW (power in the microring). The optical spectrum is shown in Fig. 2(a). (a) Temporal waveform (b) radio frequency spectrum. (c,d) Same as (a,b) for the lasing at 7mW. The optical spectrum is shown in Fig. 2(b). (e,f) Same as (a,b) for the lasing at 10mW. The optical spectrum is shown in Fig. 2(c). The DC component of the RF spectrum is indicated with a square marker.

In this regime we also observed the classic Q-switching (or gain-switching) [32] operation of the doped fiber pulsed laser, with a recovery time on the order of milliseconds, characteristic of the ytterbium gain relaxation time. In this regime the pump is characterized by unstable operation, driven by supermode beating. We note however that the bandwidth of the RF oscillations (< 20MHz 3dB bandwidth) is significantly less than the microresonator linewidth of 160MHz. While often undesired, this instability may in fact play a role in enhancing the peak power and hence improving the OPO operation, as demonstrated by the relatively low OPO threshold we achieved (around 6mW – almost an order of magnitude less than the OPO in [14]). In order to eliminate the low-frequency modulation arising from supermode instability, the same approach used [24] for the FD-FWM laser – shortening the main cavity length and properly adjusting a delay line to allow for single main cavity mode operation within each microring resonance – is also expected to work for the self-locked approach. Therefore, in the interest of realizing a fully stable complete RF spectrum for the self-locked laser, we attempted to achieve operation of the self-locked OPO using a short EDFA amplifier cavity laser similar to [24], with a main cavity FSR of 64MHz. The total power obtained by this shorter laser (around 15mW, well below the threshold in [14]), however, was not sufficient to achieve oscillation. Nonetheless, when filtering out the pump, we were able to detect the generated signal by means of a high sensitivity photomultiplier detector. By increasing the gain of the short-length amplifier we expect that this scheme should lead to stable OPO operation [24, 26]. Finally, we note that the overall power output of our OFC is comparable to other results based on microcavity frequency combs [15, 16]. We note though that our output power could easily be increased by as much as 20dB by simply using a fiber Bragg grating (FBG) centered around the pump wavelength, as a filter to extract the frequency comb lines rather than the 1% tap filter used here.

## 5. Conclusions

We demonstrate a novel self-locked pumping scheme to achieve robust operation of OPOs based on microcavities, using a nested cavity approach with a four-port micro-resonator within an amplifying fiber loop. By properly adjusting a bandpass filter, this scheme resulted in a robust oscillation intrinsically locked to the microcavity resonances and resilient to thermal or mechanical perturbations. No evidence of self-termination was observed, as is typical for externally pumped OPOs. Further, our method consistently produced very wide bandwidth (>300nm) frequency combs that also had a spacing given by the FSR of the microcavity, independent of the pump laser wavelength within the gain bandwidth of our amplifier (30nm). We achieve this in a CMOS compatible high-Q microring resonator, but our scheme can be in general applied to a large class of microresonator with different physical properties, e.g. FSR. This scheme has the potential to solve a key challenge, i.e. the sensitivity to thermal and mechanical perturbations, that has plagued externally pumped microcavity OPOs, in order to achieve a practical chip based optical frequency comb source for many applications.

**Acknowledgments**

This work was supported by the Natural Sciences and Engineering Research Council of Canada (NSERC) and the Australian Research Council (ARC) Discovery Projects and Centres of Excellence programs. L.C. acknowledges the support from the Government of Canada through the PDRF program and "*Le Fonds québécois de la recherche sur la nature et les technologies*" (FQRNT) through the MELS fellowship program.